\documentclass[useAMS,usenatbib,referee]{mn2e}
\usepackage{epsfig}
\newif\ifAMStwofonts

\newcommand{\be}{\begin{equation}}
\newcommand{\ee}{\end{equation}}
\newcommand{\bea}{\begin{eqnarray}}
\newcommand{\eea}{\end{eqnarray}}
\newcommand{\Bea}{\begin{eqnarray*}}
\newcommand{\Eea}{\end{eqnarray*}}

\title[The outer layers of Procyon A: comparison with the Sun]
{Simulating the outer layers of Procyon A: a comparison with the Sun} 
\author[Robinson et al.]
       {F.J. Robinson$^1$\thanks{Email: marjf@astro.yale.edu},
            P. Demarque$^2$, 
\newauthor
D.B. Guenther$^3$,
Y.-C. Kim$^4$\thanks{Email: kim@galaxy.yonsei.ac.kr},
K.L. Chan$^5$  \\
$^1$Department of Geology and Geophysics, Yale University, Box 208109, New Haven, CT 06520-8109, USA\\
$^2$Department of Astronomy, Yale University, Box 208101, New Haven, CT 06520-8101, USA\\
$^3$Department of Astronomy and Physics, Saint Mary's University,
Halifax,
Nova Scotia B3A 4R2, Canada\\
$^4$Astronomy Department, Yonsei University, Seoul, Korea\\
$^5$Hong Kong University of Science \& Technology, Hong Kong, China}

\begin{document}
\maketitle

\label{firstpage}

\begin{abstract}
We compare a new 3D radiation-hydrodynamical simulation of
the surface layers of Procyon A
to a similar 3D simulation of the
surface layers of the Sun.  Both simulations include realistic
input physics and are performed using the same numerical techniques
and computer codes.

Convection in the surface layers of Procyon A is very different from the Sun.
Compared to the Sun, the atmospheric structure and
 convective flow in Procyon A exhibit
the following characteristics:
(1) the highly superadiabatic transition layer (SAL) is
located at much shallower
optical depth; it is in a dynamically active region,
 and its outer region is located part of the time in
the optically thin atmosphere;
(2) the outer region of the SAL moves from an optically thin region 
to thick region and back again over a time  of 
20-30 minutes. This motion, which is driven by the granulation, takes place 
in a time approximately half the turnover time of the largest granules;
(3) the peak root mean square velocity in the vertical direction is much larger in Procyon A.

The main reason for the radically different radiative-convective behaviour  in Procyon A 
compared to the Sun is  the role played  by turbulent eddies  
in determining  the overall flow/thermal structure. 
The turbulent  pressure and turbulent kinetic energy 
can exceed  50 \%  of the local gas pressure (compared to about 10-20 \% in the Sun). 
In  such regions, the mixing lengthy theory (MLT) is a poor approximation.

The Procyon A simulation thus reveals two distinct timescales - the autocorrelation time
of the vertical velocity and the characteristic timescale of the SAL which is tied to
granulation.
Just below
the surface the autocorrelation decay time is about 5 minutes in Procyon A, and
the SAL motion timescale is 20-30 mins.
In the  simulations the 
peak value of the superadiabaticity 
varies between 0.5 and 3. When the SAL penetrates the optically thin region there are  
efficient radiative losses and the peak of the SAL is low.
We speculate that these losses damp out 
the relative amplitudes in luminosity (temperature fluctuations) compared
to velocity (Doppler). Although this will not affect the frequencies of
the peaks in the power spectrum, it will probably lower the average amplitude of the peaks
relative to the noise background.
\end{abstract}

\section{Introduction}

Procyon A (hereafter referred to Procyon),
one of the brightest stars in the sky, is a well
studied object.  Known for its white dwarf companion,  its
orbit has been well determined by speckle technique, and its
mass, long quite uncertain, has now been evaluated to be near 1.5 $M_{\odot}$
(Girard et al. 2000).  
On the basis of evolutionary interior models, Procyon's internal structure
is believed to consist of a convective core,
with a radiative envelope and a thin surface convection zone. The mass 
in its outer convection zone is 
$10^{-5} M_{\odot}$ at present (Guenther \& Demarque  1993).  
Procyon thus offers an unusual opportunity to study 
by seismic means both the structure of the outer layers 
and the efficiency of convective core
overshoot in intermediate mass stars (Chaboyer et al. 1999; Straka et al. 2005).

For all these reasons, and in view of theoretical predictions that the
$p$-mode oscillation amplitudes would be relatively large
(Christensen-Dalsgaard \& Frandsen 1983; Houdek et al. 1999), Procyon has been a prime 
candidate for seismic  
observations from the ground and in space.
Interestingly, now that reliable observations are possible,
there is an apparent discrepancy  between the detection of
$p$-mode
oscillations on Procyon, based on radial velocity measurements 
made from the ground (Marti\'{c} et al. 2004 and Eggenberger et al. 2004),
and the high precision intensity measurements
described by the MOST space mission (Walker et al. 2003).  In contrast with the
ground based observations, the MOST space telescope has failed  to detect
$p$-modes from Procyon (Matthews et al. 2004) at its instrumental limit.

This paper focusses on modeling the dynamics and physical conditions in the
outer convective layers of Procyon, and in making a comparison  with the Sun.
We present the results of a 3D radiative hydrodynamic
numerical large eddy simulation of the outer layers of Procyon, and
compare it to a similar simulation performed previously for the Sun using the
same detailed physics input and numerical code
(Robinson et al. 2003).
The domains of both simulations
extend from the observable atmospheric layers in radiative equilibrium and
the outer layers of the convection zone,
down to a depth at which the temperature gradient exceeds the adiabatic
value by only a very small amount (deep convection). The simulation  domain
in both simulations includes the highly superadiabatic transition layer
(hereafter SAL) between the atmosphere and deep convection.
It is in the SAL region that the $p$-modes
are believed to be excited by stochastic processes.

In the next section, we describe briefly the starting model and the
procedure followed in
carrying out the 3D numerical simulation.
Section~3
discusses the main results from the Procyon simulation, emphasizing the
striking differences in the physical conditions in the atmospheres of
Procyon and the Sun.
The most surprising result is that the
SAL, which in the Sun changes little with time and is
placed below the photosphere,
is found to vary with time in position and amplitude in Procyon.
As a result, the outer layers of the SAL in Procyon are found 
part of the time in the optically thin atmosphere in Procyon. In this region,
radiative losses are large, and characteristic
timescales are short, in contrast
with the Sun where the SAL is subphotospheric.

The implications of these
fundamental differences for the excitation, damping and lifetimes of $p$-mode
oscillations, are discussed in general terms in Section~4, 
where we analyze the likely consequences for $p$-mode observations
in Procyon, and the conditions for the detectability of $p$-mode
oscillations either
from the intensity measurements from space (MOST mission),
or from ground-based
radial velocity measurements.

In Section~5, we summarize our results and the direction of future 
research in the light of our study.

\section{Three-dimensional simulation}

The 3D numerical simulation of Procyon was carried out using
the same numerical approach and physical assumptions as in our
simulations of the outer layers of the Sun
(Robinson et al. 2003) and subgiant stars (Robinson et al. 2004).
As in the case of the Sun, the same description of the  
microphysics, i.e. the opacities and equation of state,
was assumed in the 3D simulation as in the 1D starting model described below. 
At low temperatures,
we used the Alexander et al. (1994) opacities, and the OPAL opacities at higher
temperatues (Iglesias \& Rogers 1996). The equation of state
was the OPAL equation of state (Rogers, Swenson \& Iglesias 1996).

\subsection{Starting 1D model}
The  first step was to construct a detailed one-dimensional (1D)
evolutionary model for Procyon, with the help 
of the Yale stellar evolution code
YREC (Guenther et al. 1992; Guenther \& Demarque 1997).
This model then provided the starting model for the 3D simulation. 

The choice of chemical composition for the simulation is of some interest.  
Evolutionary models of Procyon show that the mass in the
surface convection zone  is
very small, varying from $10^{-7} M_{\odot}$ at minimum,
to $10^{-5} M_{\odot}$ at present
(Guenther \& Demarque  1993).  As a consequence,
gravitational settling is expected to take place below the convection zone of Procyon, raising 
questions about the chemical composition to use in modeling the atmosphere of 
Procyon.

Heavy element diffusion in the envelope 
is complicated by the fact that
both radiative levitation of certain elements and turbulent mixing may
also play an important role in inhibiting the
gravitational settling (Richer et al. 1998). The Procyon atmosphere
typifies the complex transition between the
cooler sun-like stars with a deep convection zone and hotter stars
along the main sequence where radiation dominates in the envelope.
From the point of view of this simulation, however, which considered only the 
turbulent outermost layers, we could safely ignore these refinements.   
Procyon's metallicity has been measured
spectroscopically to be close to solar (Takeda et al. 1996; Kato,
Watanabe \& Sadakane 1996).  For simplicity, we adopted
the same metallicity mixture as for the Sun (Grevesse \& Noels 1993).

The situation is more uncertain regarding the helium abundance because the 
helium abundance of Procyon cannot be determined spectroscopically.
In fact, in the absence of any other effect, applying the treatment of 
gravitational settling which is satisfactory for the Sun, 
would result in the complete depletion
of helium in the Procyon atmosphere in less
than $10^8$ years (Guenther \& Demarque 1993; Provost et al. 2005).
It has been suggested by Morel \& Th\'{e}venin (2002) that the settling efficiency
is reduced greatly by radiative diffusivity effects. 
In view of the above uncertainties, the starting model was constructed under 
the arbitrary assumption that the efficiency of gravitational settling is  
decreased by two orders of magnitude.  This resulted in a helium content 
by mass of Y = 0.07.  
We emphasize that
the precise value of Y is of little consequence in this study of the
structure and dynamics of the atmosphere. In stellar atmospheres of 
this temperature 
helium is ``dead weight'', and variations in helium content can 
be mimicked by 
a small adjustment of the effective gravity.  Thus in contrast to the case 
of the heavy elements, which affect the radiative opacities sensitively, 
the precise 
helium abundance is a second order effect in determining the structure and 
dynamics of the atmosphere.    

\subsection{The 3D simulation}
\begin{figure}
\epsfig{file=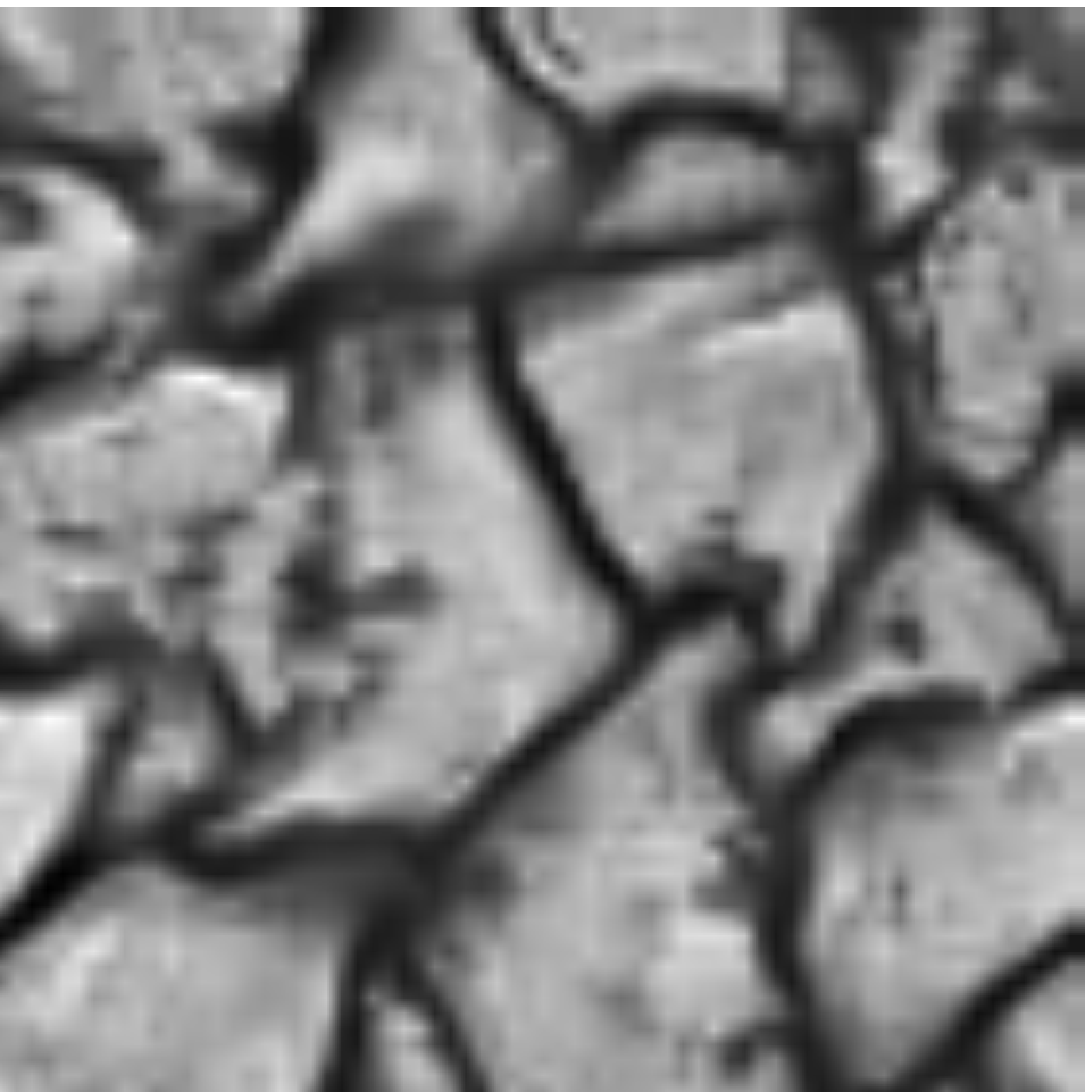,width=8cm}
\caption{Instantaneous vertical velocity in a horizontal plane located in the SAL of Procyon.
Dark regions denote downflows, lighter regions are upflows.
The dimensions are $29000 {\rm km} \times 29000 {\rm km}$}
\label{29Mm-gran}
\end{figure}

The 3D simulation of Procyon is a square based box of  dimensions 14500 km $\times$    14500 km  $\times$  16300 km  
on a $74 \times 74 \times 160 $ uniformly spaced grid.  The vertical extent is about 
5.5 pressure scale heights which includes the helium and  hydrogen ionisation zones and 
part of the overlying  radiation region.
As the simulated granules have diameters of about 10000 km one could argue that a 
larger domain may be required. However, the effect of 
the box width on the thermal structure or turbulent pressure 
was shown to be minimal (Robinson et al. 2003), though it may have a small effect on the 
turbulent kinetic energy near the top. As the focus our study is on  
the structure of the SAL and the vertical velocities,  the box width should be adequate. 
The importance of the depth and width of the box is discussed in detail in 
Appendix A of Robinson et al. (2003).   
To make sure that the box is indeed big enough, we ran an additional simulation
in a box of dimensions 29000  km $\times$   29000 km  $\times$  16300 km.
A grey scale plot of the instantaneous vertical velocity at a horizontal cross-section
located near the peak of the SAL is shown in Fig.~\ref{29Mm-gran}  The size and shape of the granules
does not appear to be affected by the side boundaries of the domain.
We confirmed that the average thermal structure and root mean square velocities (not shown) are almost the same on the
29000  km wide box as they are on the  14500 km  wide box.

The 3D simulation of the Sun is case D in Robinson et al. (2003). This 
has dimensions 2700 km  $\times$ 2700 km  $\times$ 2800 km 
on a $58 \times 58 \times 170$   grid. The Sun simulation does not extend down  to the 
helium ionisation zone but still contains  about 8.5  pressure scale heights. 

\section{Comparison with the Sun}

\subsection{Mean atmospheric structure and superadiabatic layer}

\begin{figure}
\epsfig{file=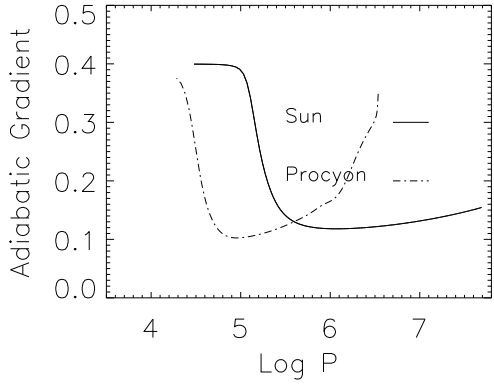,width=8cm}
\caption{Mean adiabatic gradient in the Sun and Procyon.}
\label{nad}
\end{figure}

We begin by comparing the basic structural parameters of the
atmospheres of the Sun and Procyon.
Unless otherwise specified, given quantities will be  temporal and horizontal averages computed after the 
simulation has reached a statistically steady state (each horizontal average is take at fixed height).
Fig.~\ref{nad}  shows the mean adiabatic logarithmic gradient $\nabla_{\rm ad}$
as a function of depth
(expressed in terms of log P), within the two atmospheric simulations.
The quantity $\nabla$, defined in the usual way, i.e.:

\begin{equation}
\nabla  \equiv  \frac{d \ln T}{d \ln P}
\end{equation}

is a statistical quantity in the simulation.
Partial ionisation reduces the value of $\nabla_{\rm ad}$ from the monatomic gas
value of 0.4, as shown in Fig.~\ref{nad}.
We note that in the case of Procyon, both the hydrogen and helium ionisation zones
are included in the simulation domain.  In the case of the Sun,
only part of the hydrogen ionisation region is included in the simulation domain.

Both Procyon and the Sun exhibit a surface convection zone which is
nearly isentropic in the deep layers.
Near the surface, the transition between deep nearly adiabatic convection
to the radiative atmosphere is characterized by the narrow highly superadiabatic
layer, denoted above as SAL.  In this layer, the specific entropy rises steeply.  This
entropy jump is known to be critical in specifying the internal structure of
the stellar model,
in particular the depth of the
convection zone and the model radius, which depend on the specific entropy
in the deep adiabatic region.  It also affects
sensitively the local sound speed near the surface
and the observed $p$-mode frequencies.
Most importantly, the SAL is the region where the balance between stochastic
excitation from turbulent motions and competing radiative damping
takes place, thus determining the amplitudes and lifetimes of $p$-modes.

\begin{figure}
\epsfig{file=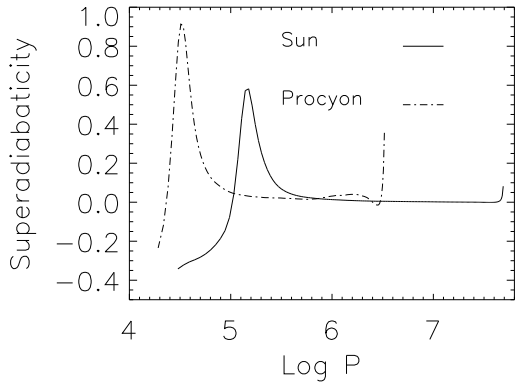,width=8cm}
\caption{Mean superadiabaticity in the Sun and Procyon. The vertical lines indicate the  approximate location of the photosphere.}
\label{sal}
\end{figure}

The excess of the 
temperature gradient over the adiabatic gradient, or superadiabaticity, is given 
by:
\begin{equation}
\nabla-\nabla_{\rm ad} = {\frac{d \ln T}{d \ln P}}- {\nabla_{\rm ad}}
\end{equation}
where the subscript ``ad'' refers to the adiabatic value.  
Fig.~\ref{sal}  shows the region where the mean superadiabaticity peaks as 
a function of log P in the Sun and in Procyon.

A major difference between the two simulations is that the peak 
of the SAL in Procyon  
is in a region of much lower density than in the Sun.  The SAL 
peak  in the Sun is located below the 
photosphere in a region that is relatively optically thick, and although 
the solar photosphere is affected by the convective flows just below, the 
position of the photospheric surface suffers little change due to the 
convective motions.  This is in sharp contrast with Procyon, where the 
more violent convective flows are present in the atmosphere and cause  
the SAL position to vary with time. The SAL moves  radially back and forth  
over a time of about 20-30 minutes and over a distance of up to  about half a local pressure scale height 
(about  500 km). 

By computing the Full Width at Half-Maximum   
of the vertical velocity we obtain  a  mean granule diameter of about  10000km (details of 
how  we estimate the granule diameter are given in Robinson et al. (2004), 
while the  run of rms vertical velocity,  Fig. \ref{vzrms}  provides a velocity scale 
for the granules of 5-7 km/s.  
Combining these quantities yields an average time  scale, roughly half the time it 
takes the largest granules to overturn fully, 
in the vicinity of 20-30  minutes.
A 42 minute movie of granulation in Procyon
for the box  is presented at  $ www.astro.yale.edu/marjf $. The largest granules (seen in the
top right hand corner of the picture) last about 50 minutes.

The overturning granular motion causes the  SAL peak to approach 
the optically thin layers part of the time.
 At one point in the simulation when the SAL
is in a particularly shallow region the  very large radiative losses produce a
downflowing plume that is able to traverse the entire vertical extent of the computational domain.
This is shown clearly in a visualisation of the convecting plumes in Procyon
given at  $ www.astro.yale.edu/marjf $

\begin{figure}
\epsfig{file=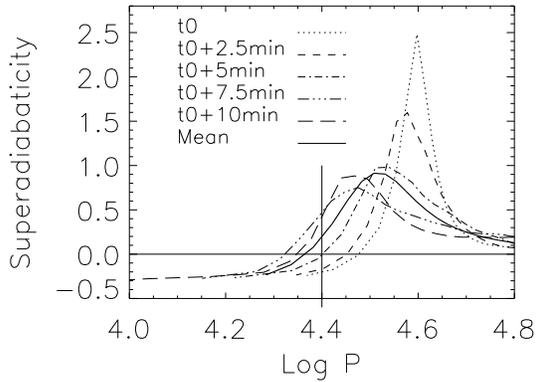,width=8cm}
\caption{Instantaneous superadiabaticity (horizontally averaged) at 2.5 minute intervals. The time average over the entire simulation is 
also plotted for reference. The solid vertical line approximatley indicates the mean location of the photosphere.}
\label{sal_procyon}
\end{figure}

\begin{figure}
\epsfig{file=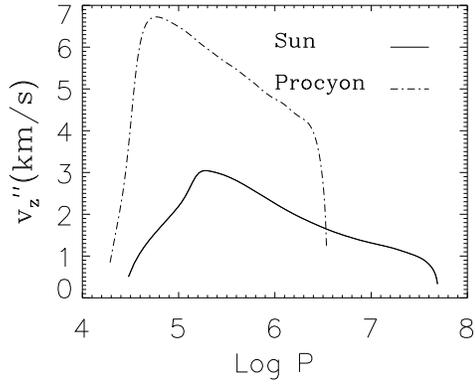,width=8cm}
\caption{Root mean square vertical velocity in the Sun and Procyon.}
\label{vzrms}
\end{figure}

To illustrate the response of the   SAL to the 
turbulent overturning granules we plot the superadiabaticity at 5 instants  
in Fig.\ref{sal_procyon}. The photosphere, defined by the 
location at which the horizontally and temporally  averaged temperature 
equals the effective temperature of the 1D Procyon stellar model, is 
marked by a solid vertical line.  The solid horizontal line shows where the  superadiabaticity is 
zero.
Each plot is separated by  2.5 minutes in time. 
The time 
$t0$ is chosen as the time after which  
the system has reached statistical thermal equilibrium.
The figure shows that over the 'quasi-periodic' cycle the SAL varies  both in  position and  height.
During part of the cycle  the outer region of the 
SAL lies above the photosphere.

By sampling at various instants we found that this `quasi-periodic' radial movement of the SAL is
maintained  
over at least  250 minutes  (the entire statistical integration time).
We note in addition that at certain times the superadiabatic peak in Procyon reaches 
twice the height of 
the solar superadiabatic peak. 
When the outer part of the 
SAL is in an optically thin region, the photons can more readily carry away
excess internal energy from the temperature fluctuations, 
than they can in optically thick regions. 
As it looses its excess internal energy compared to its
surroundings at a greater rate, the
fluctuation is said to be radiatively damped. 
Because of this damping it is reasonable to infer that the $p$-mode
intensity amplitudes (which depend on the temperature fluctuation) 
will be smaller  when the SAL is in optically thinner regions.

\subsection{Velocity fields, turbulent pressure and turbulent kinetic energy}

The root mean square  vertical velocity is plotted as a function 
of log P in both the Sun and Procyon in Fig.\ref{vzrms}.  
We see that the turbulent
velocity has a much larger amplitudes in Procyon than in the Sun.
The Procyon simulation predicts vertical velocities
and velocity fluctuations in the optically thin atmosphere 2-3 times larger than in the Sun.
This is consistent with observation. A recent spectroscopic study of Procyon
by Allende Prieto et al. (2002) concludes that a comparison of the velocity spans
(in line bisectors) for the
Sun and Procyon shows that ``the span of Procyon's lines exceeds the solar values by more
than a factor of 2''.

A gray scale map  of vertical velocities near optical depth unity, as previously shown in Fig. \ref{29Mm-gran},
display a pattern of granulation in Procyon similar to that observed in the Sun,
but more chaotic and on a larger scale.
The lighter regions denote upflows, while the darker regions at
granule boundaries denote downflows. There is a difference of scale; in
Procyon, the granule sizes average about  10,000 km, while solar granules
average  about 1200 km in horizontal scale.  

\begin{figure}
\epsfig{file=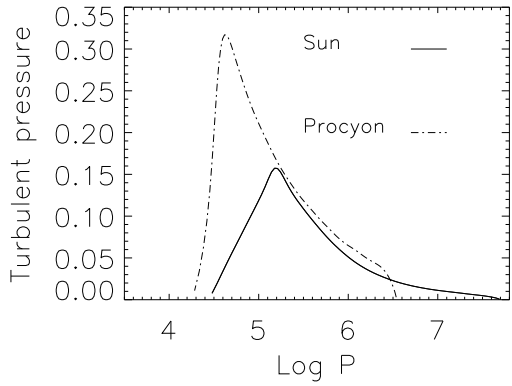,width=8cm}
\caption{Turbulent pressure divided by the local gas pressure in the Sun and Procyon.}
\label{pturb}
\end{figure}

A useful quantity is the ratio $R$ of turbulent pressure 
$P_{turb}$ to mean gas pressure ${\overline P}$. The 
overbar denotes a combined horizontal and temporal average. 
This ratio can be written: 
\begin{equation}
{R = P_{turb}/{\overline P}}= {{\overline \rho}} {v_z''}^2 /{\overline P},
\end{equation}
which is a non-dimensional quantity.
The quantities $\rho$ and $v_z''$ are the average  density and root mean square vertical 
velocity, defined as 
\be
{v_z''}^2={\overline{ {v_z}^2}}\hspace{1mm} - {\overline {v_z}}^2
\ee
where again the overbar denotes a combined horizontal and temporal average.
Fig. \ref{pturb}  is a plot of   $R$ 
as a function of depth for Procyon and the Sun. 
Unlike the case of the Sun, in Procyon the turbulent pressure varies `quasi-periodically' with the 
position of the SAL  and at times the  instantaneous horizontal average 
of the turbulent pressure divided by the gas pressure can have  a peak value of as much as 50 \%.   The 
instantaneous horizontally averaged turbulent kinetic energy divided by the local gas pressure can also reach 
50 \% near the top of the box (not shown).
In the simulation the maximum 
Mach number 
(defined as the maximum $v_z''$ divided by the local sound speed) is about 2 near the top 
(in the 
vicinity of the SAL peak).
So motions will be supersonic at small optical depth suggesting short characteristic
time scales.

\subsection{Overshoot}

\begin{figure}
\epsfig{file=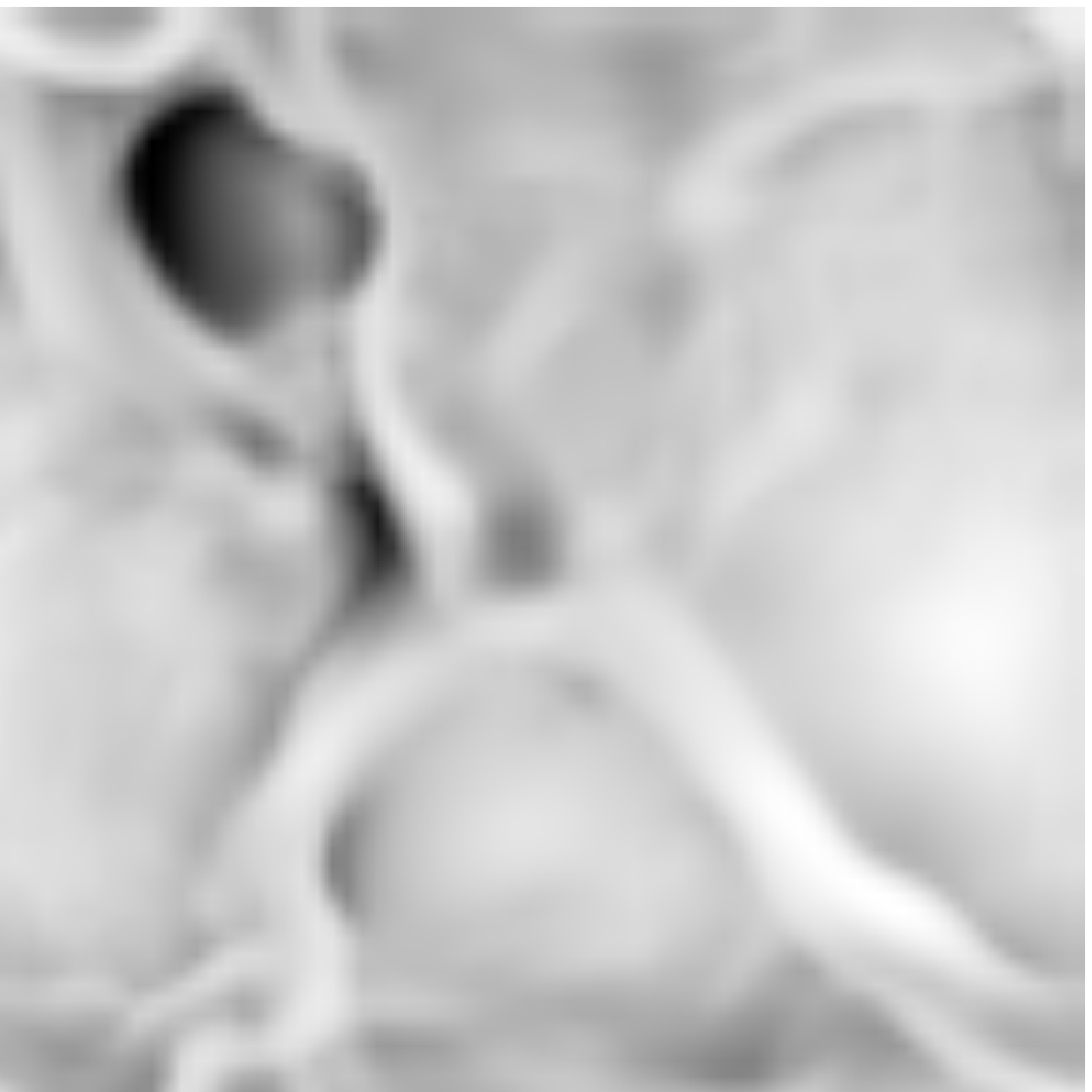,width=6cm}
\epsfig{file=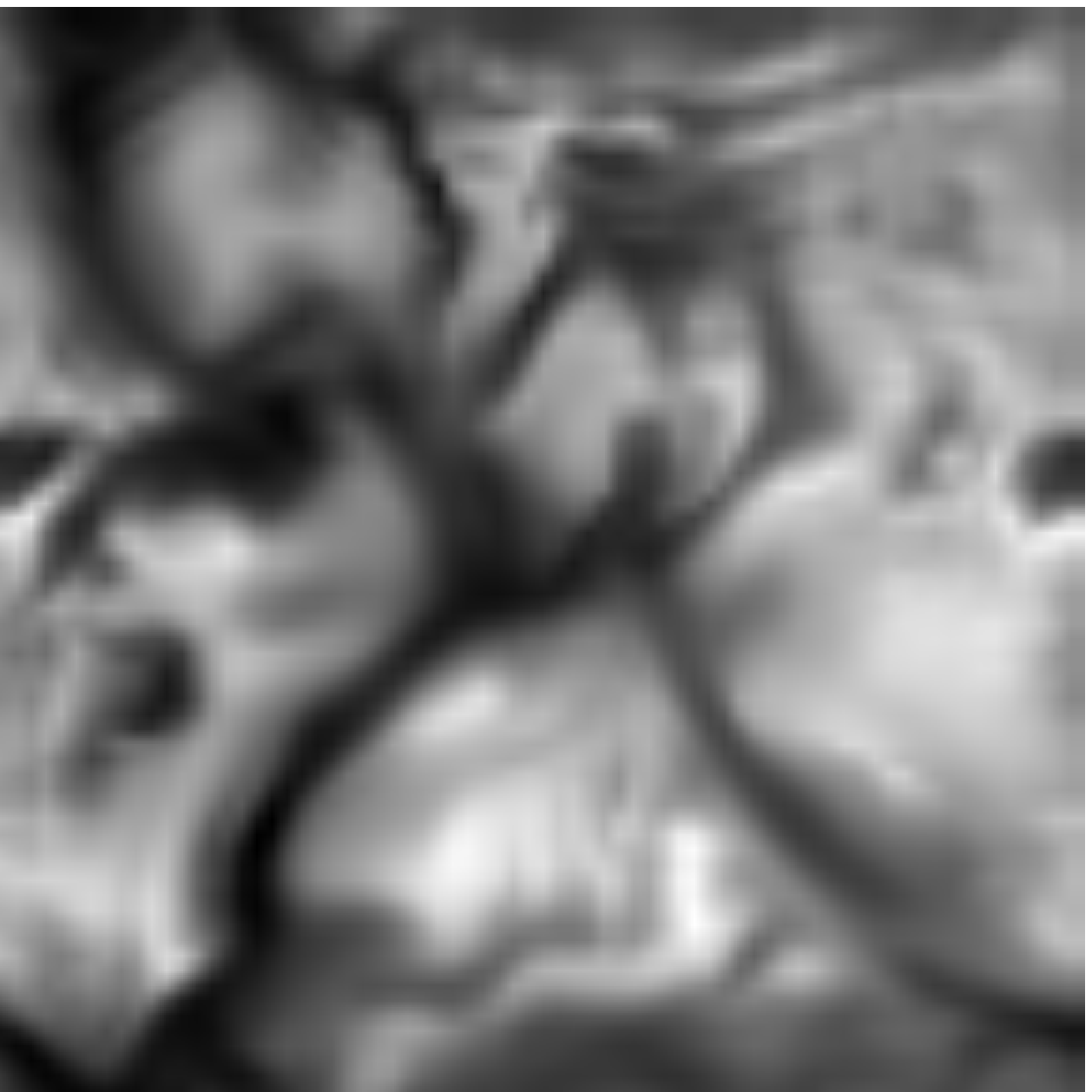,width=6cm}
\caption{The {\it left} panel is a  gray scale plot of the temperature fluctuation
in a horizontal plane located in the 
overshoot region just above the photosphere of Procyon. 
Lighter regions
indicate hotter than average fluid while darker regions
indicate cooler than average fluid.
The {\it  right} panel is the corresponding vertical velocity 
at the same instant but located just inside  the photosphere. 
Dark regions denote downflows, lighter regions are upflows.
The dimensions are $14500 {\rm km} \times 14500 {\rm km}$}
\label{vzcontour}
\end{figure}

By comparing 
the horizontal cross-section of the 
temperature fluctuation  (left panel of Fig.\ref{vzcontour}) at shallow optical depth ($\tau$ = 0.001)
with the  vertical velocity at optical depth unity just inside the photosphere (right panel of Fig. \ref{vzcontour}),
one can identify regions in which hotter (colder) than average fluid 
moves down (up).  At small optical depth, the downflows appear brighter (hotter) 
(instead of darker below the photosphere) because the fluid is being adiabatically compressed.  
This is a signature of convective $\it{overshoot}$ above the photosphere. 
Note  the dark (cold) spot in the upper left hand corner of  the left panel. This 
indicates  upward moving  fluid  that is colder than the horizontal mean temperature. 
Generally only a few  such  updrafts are sufficiently energetic  to  be able 
to continue travelling up even though they are negatively buoyant. 

This reversal of the granulation pattern (bright downflows and darker 
granular upflows) ``at large height''  was also noted noted by Nordlund \& Dravins (1990) in 
their study of the Procyon atmospheric granulation.  

\subsection{Comparison with earlier studies}
Our main conclusions regarding the time dependence of granulation and
of the SAL in Procyon also appear robust.
It is interesting to compare the results of our simulation
to an earlier less detailed hydrodynamic study of Procyon's atmospheric structure
and granulation (Nordlund \& Dravins 1990; Dravins \& Nordlund 1990).
These authors included time dependent hydrodynamics in their non-grey
atmosphere calculations.  Even though the model were relatively
coarser than ours (the simulation volume included $32 \times 32 \times 32$ elements),
and a simplified treatment of the hydrodynamics (the anelastic
approximation) was used,
the larger temperature contrast of the visible surface was present. Nordlund 
\& Dravins (1990) also noted the position of the granules in the thin
atmosphere (``naked'' granules), in contrast with the Sun.
Some features of the striking difference
between the properties of
convection in an F-type main sequence stellar atmosphere and the solar atmosphere had
already been noted by Nelson (1980) using a simpler model of granulation
developed for the Sun (Nelson \& Musman 1977).

\section{Excitation depth of $p$-modes in the Sun and Procyon}

Stochastic fluctuations in the turbulent layers
are believed to be responsible for the excitation of $p$-modes in the
Sun and stars with similar convection zones.
The 3D numerical simulations of the solar atmosphere have
been successful in explaining the excitation mechanism and
power distribution of solar $p$-modes (Stein \& Nordlund 2000).
We have seen that Procyon exhibits a much shallower convection zone
than the Sun, and it is interesting to query whether the
SAL region in Procyon also favors the excitation of $p$-modes.

The turbulent velocity fluctuations in a convection zone must be 
stochastically distributed for the excitation of $p$-modes to take place.
We suggest that  only when the coherence time for the velocity fluctuations becomes 
shorter than the period of the oscillation modes where the oscillation 
power is concentrated, can the fluctuations contribute to the stochastic excitation.
To enable stochastic (random motions) excitation of $p$-modes by turbulence,
the turbulent velocity (or temperature)  field should be stochastic over a time interval  smaller than
the $p$-mode oscillation it is intended to excite. Otherwise its not temporally stochastic with
respect to the  oscillation.
For the Sun, the relevant oscillation period is  $\wp\sim 5$ minutes, 
in Procyon, it is believed to peak around $\wp\sim 15$ minutes 
(Marti\'{c} et al. 2004). 

The autocorrelation coefficient of vertical velocity (which is a function of
depth and time)
is defined as
\be
C[v_z'(0)v_z'(t)]=\frac{\langle{v_z(0)v_z(t)}\rangle-
\langle{v_z(0)}\rangle\hspace{1mm}\langle{v_z(t)}\rangle}{v_z''(0)v_z''(t)}
\ee
where
\be
v_z''=\sqrt{\langle{v_z}^2\rangle-{\langle{v_z}\rangle}^2}
\ee
and $v_z'=v_z-{\langle{v_z}\rangle}$.
The angled brackets represent a horizontal average.
For brevity of notation, we have written
$v_z(x,y,z,0)$ as $v_z(0)$, and
$v_z(x,y,z,t)$  as $v_z(t)$.

\begin{figure}
\epsfig{file=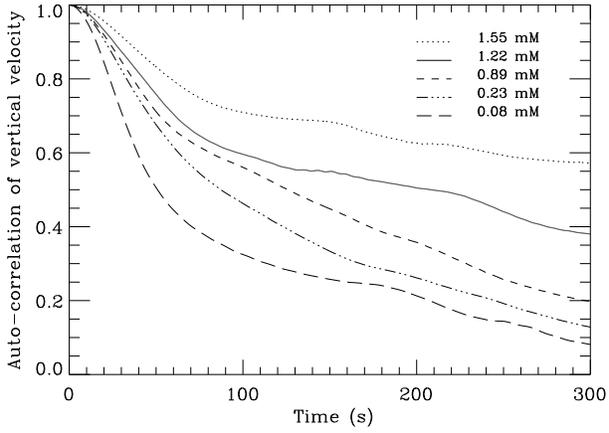,width=8cm}
\caption{Auto-correlation of vertical velocity at various depths with respect to
the photosphere in the Sun. Depth is measured positively inwards from the
photospheric surface. After about 5 minutes the
auto-correlation coefficient 
is about 0.1 at a depth of about
0.1Mm below the photosphere.}
\label{corr_vz_sun}
\end{figure}

Figure \ref{corr_vz_sun} depicts the auto-correlation
coefficient of
vertical velocity fluctuations, as a function of time and depth,
from the 3D numerical simulation of
the near-surface layer of the convection zone in the Sun. The
autocorrelation coefficient varies between 0 and 1. The
vertical velocity fluctuation can be considered
to be stochastic when the autocorrelation coefficient becomes small, say
of order of 0.1.
In the Sun, we find that after 5 minutes, the
auto-correlation coefficient decreases below
0.1 at a depth of about
0.1Mm below the photosphere.  This corresponds to the
depth where $\log P\approx 5$.
This is near the SAL peak,
just below the photosphere, and a depth favorable to $p$-mode excitation.

\begin{figure}
\epsfig{file=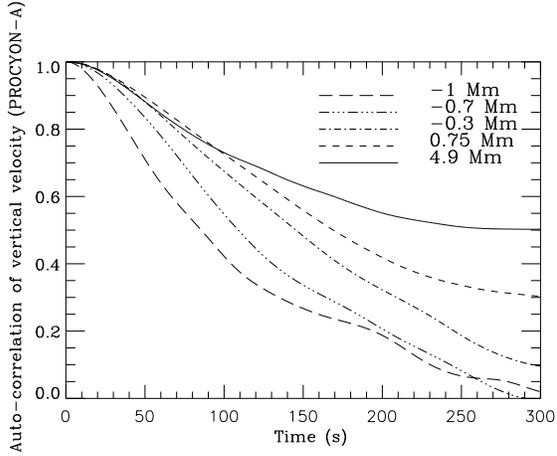,width=8cm}
\caption{Auto-correlation of vertical velocity at various depths with respect to 
the photosphere in Procyon. Depth is measured positively inwards from the 
photospheric surface. After about 5 minutes the
auto-correlation coefficient
near the photosphere is  about 0.1.  
As the position of the SAL varies significantly over the simulation,
the details of the plot will depend on the starting time.}
\label{corr_vz_procyon}
\end{figure}

An autocorrelation plot started at  one instant during the simulation of  Procyon is shown in Fig. \ref{corr_vz_procyon}.
We see that in the Procyon atmosphere, the vertical
velocity fluctuations become  stochastic after a time of the
order of 300 seconds, i.e. compared to the oscillation period $\wp$, motions in Procyon
are much more stochastic in the shallow layer than in the Sun.
At other times during the simulation the motions become stochastic 
in a time of between  200 and 600 seconds.  This is because of the 
radial movement of the SAL described in the previous section.
The conditions for stochastic excitation of $p$-modes would
seem to be favorable for modes with periods in the vicinity
of 15 minutes.  In fact, several observers
who made radial velocity measurements of Procyon from the ground have
found evidence for
such oscillations (Brown et al. 1991,  Marti\'{c} et al. 2004, Eggenberger et al. 2004).

We emphasize that above arguments based on the autocorrelation coefficient for 
the velocity fluctuations suggest only that stochastic 
excitation can indeed take place in the outer layers of Procyon, at least part of the time,
during the quasi-periodic SAL oscillation.  
This approach does not 
tell us whether a particular mode is excited or not by the stochastic process.  In other words,
we use it as a necessary, but not a sufficient condition for mode excitation.
Mode excitation would require a detailed calculation of the 
physical response of the convective region to the propagation of sound waves within it.
Such a calculation, while quite feasible, is outside the scope of the present paper.

It is however possible to infer on general physical grounds on the 
differences between the Sun and Procyon.  Because large temperature fluctuations due to 
convection take place in Procyon in regions of lower density, they will come into 
equilibrium with their surroundings more rapidly than in the Sun, sometimes on a timescale 
shorter that the period of the shear motions associated with the acoustic oscillations.   
This seems to be the case in the simulation, where much happens to the 
SAL position on a timescale of  say only 10 minutes (see Fig.3).


Energy losses by
radiation on short time scales would in all likelihood overwhelm low amplitude
intensity fluctuations with longer periods,
such as the $p$-modes with periods in the vicinity of 15 minutes. 
Thus the detection of intensity fluctuations due to
$p$-mode oscillations in the Procyon atmosphere would seem problematic.

The time-dependent granulation would also introduce a source of noise
in radial velocity measurements. However, in the absence of a
more detailed knowledge of the relative strength of the
excitation and damping mechanism operating in Procyon's atmosphere,
it is difficult to ascertain the precise effect on
radial velocity $p$-mode measurements.

\section{Summary and conclusions}
We have performed physically realistic numerical 3D simulations of the
outer layers of Procyon, and have compared the results with a 3D simulation
of the solar outer layers performed under the same assumptions.
The most striking difference between the Sun and Procyon appears in the
atmosphere dynamics. In Procyon, the optically thin atmosphere is
subject to quasi-periodic changes on timescales of between 20 and 30 minutes. 
This is roughly 
half the complete overturn timescale of the largest granules 
in the optically thin surface layers.
Matthews et al. (2004) present evidence for the detection of unusually 
strong granulation noise in Procyon in the MOST space mission data.
Future MOST observations of Procyon may be capable of detecting the signature of the 
quasi-periodicity in the granulation revealed by our simulation. We emphasize that 
this atmospheric phenomenon would not have a fixed period  
since the velocities and 
dimensions of the granules are stochastic.  
The presence of granulation in the upper atmosphere of
Procyon was
also pointed out by Nordlund \& Dravins (1990); they refer to the phenomenon 
as ``naked granulation''.

What can be further inferred from our simulation regarding $p$-mode oscillations in
Procyon ?
The  highly stochastic nature of the vertical velocity fluctuations in the
Procyon atmosphere appears favorable to the excitation of $p$-modes.
On the other hand, the atmosphere is characterised 
by radiation losses tending to damp out oscillations very efficiently.  
Based on our simulation results,  we speculate on simple physical grounds, that if 
excited, $p$-modes may have a short 
lifetime and smaller relative amplitudes in luminosity (temperature fluctuations) compared
to velocity in the Procyon atmosphere.

The task of determining the relative importance of these competing exciting and
damping processes
in the dynamical and inhomogeneous environment of Procyon's atmosphere
is beyond the scope of this study.  Existing theoretical work on the 
amplitudes of stellar pulsations can perhaps serve as a guide (Houdek et al. 1999).
Because it is based on a version of the MLT (Balmforth 1992), which predicts large 
convective velocities at the surface in stars like Procyon, 
the Houdek et al. theory predicts large $p$-mode oscillation amplitudes.
The meaning of relating convective velocities to oscillation amplitude is 
unclear, however, when the stochastic excitation process relates the 
oscillation amplitudes physically 
not to the convective flow velocities themselves but to the statistical 
fluctuations from the mean convective flow in the vertical direction.
 Predicting the luminosity amplitudes is 
particularly difficult due to the uncertain coupling between the oscillations and the 
radiation field. On the basis of their models, Houdek et al. (1999) predict 
amplitudes in luminosity about 20\% of the velocity amplitudes for the Sun, which is in  
reasonable agreement with observation (Schrijver et al. 1991; Elsworth et al. 1993).  
Calculating the ratio of velocity to intensity amplitudes in Procyon 
is a major challenge ahead.

Further observations of Procyon with the MOST instrument 
and improved data from ground based obervatories will be needed to 
guide future theoretical progress.  
It is at this point reasonable to conclude that
because of the large intensity contrasts and small
time scale variations taking place in the Procyon atmosphere,
detecting the minute intensity fluctuations expected from $p$-mode oscillations,
would be a daunting task.  This difficulty, in addition to the likelihood of  
short $p$-mode lifetimes for Procyon, is a plausible explanation for the inability 
of the MOST space mission to detect $p$-modes oscillations in
Procyon (Matthews et al. 2004).

\section*{Acknowledgments}
 
This research was supported in part by the NASA EOS/IDS Program (FJR),
and by NASA grant NAG5-13299 (PD).  DBG acknowledges
support from an operating research grant from NSERC of Canada.
YCK was supported by a Korea Research Foundation grant KRF-2003-015-C00249.
KLC acknowleges support from a grant of the Hong Kong RGC (HKUST6119/02P).


\end{document}